\documentclass[twocolumn]{aa} 
\usepackage{graphicx}
\usepackage{txfonts}
\usepackage{natbib}
\usepackage{lscape}

\def\champp {CHAMP$^+$}
\def\micron {$\mu$m}

\def\Lsol {L$_\odot$}
\def\Msol {M$_\odot$}
\def\HII  {H{\sc ii}}
\def\HH {H$_{2}$}

\def\CII  {C{\sc ii}}
\def\OI   {O{\sc i}}

\def\TWCO {$^{12}$CO}
\def\THCO {$^{13}$CO}
\def\CXVIIIO {C$^{18}$O}

\def\kms {km~s$^{-1}$}
\def\vlsr {$V_{\rm LSR}$}

\def\aap{A\&A}                
\def\aapr{A\&A~Rev.}          
\def\aj{AJ}                   
\def\apj{ApJ}                 
\def\apjl{ApJL}                
\def\apjs{ApJS}               
\def\araa{ARA\&A}             
\def\nat{Nature}              
\begin{document}
\title{The APEX-{\champp} view of the Orion Molecular Cloud 1 core}

   \subtitle{Constraining the excitation with submillimeter CO multi-line observations}


\offprints{T.-C. Peng}


   \author{T.-C. Peng\inst{1,2,3}
          \and F. Wyrowski\inst{1}
          \and L. A. Zapata\inst{4}
          \and R. G{\"u}sten\inst{1}
          \and K. M. Menten\inst{1}
          }

   \institute{Max-Planck-Institut f\"ur Radioastronomie (MPIfR),
              Auf dem H\"ugel 69, 53121 Bonn, Germany 
\and Universit\'{e} de Bordeaux, Observatoire Aquitain des Sciences de l'Univers, 2 rue de l'Observatoire, BP 89, F-33271 Floirac Cedex, France 
\and CNRS, UMR 5804, Laboratoire d'Astrophysique de Bordeaux, 2 rue de l'Observatoire, BP 89, F-33271 Floirac Cedex, France 
              \email{Tzu-Cheng.Peng@obs.u-bordeaux1.fr}
              \and Centro de Radioastronom\'\i a y Astrof\'\i sica,
Universidad Nacional Aut\'onoma de M\'exico, Morelia 58090, M\'exico
             }

\titlerunning{The APEX-{\champp} view of the OMC-1 core}
\authorrunning{Peng et al.}

 
  \abstract
   {}
   {A high density portion of the Orion Molecular Cloud 1 (OMC-1) contains the prominent, warm Kleinmann-Low (KL) nebula  
 plus a farther region in which intermediate to high mass stars are forming. Its outside is affected by ultraviolet radiation from the neighboring Orion Nebula Cluster and forms the archetypical photon-dominated region (PDR) with the prominent bar feature. Its nearness makes the OMC-1 core region a touchstone for research on the dense molecular interstellar medium and PDRs.} 
    {Using the Atacama Pathfinder Experiment telescope (APEX), we have imaged the line emission from the multiple transitions of several carbon monoxide (CO) isotopologues over the OMC-1 core region. Our observations employed the $2\times7$ pixel submillimeter \champp\ array to produce maps ($\sim300\arcsec\times350\arcsec$) of \TWCO, \THCO, and \CXVIIIO\ from mid-$J$ transitions ($J=6-5$ to $8-7$). We also obtained the \THCO\ and \CXVIIIO\ $J=3-2$ images toward this region.}
      {The \TWCO\ line emission shows a well-defined structure which is shaped and excited by a variety of phenomena, including the energetic photons from hot, massive stars in the nearby Orion Nebula's central Trapezium cluster, active high- and intermediate-mass star formation, and a past energetic event that excites the KL nebula. Radiative transfer modeling of the various isotopologic CO lines implies typical \HH\ densities in the OMC-1 core region of $\sim 10^{4}-10^{6}$ cm$^{-3}$ and generally elevated temperatures ($\sim 50-250$ K). We estimate a warm gas mass in the OMC-1 core region of 86--285 \Msol.}
   {}


   \maketitle
%

\section{Introduction}


\begin{table*}
\centering
\caption{\label{table1}Observational parameters}
\begin{tabular}{lcrrcccr}
 \hline \hline

Molecule/Line & Frequency\tablefootmark{a} & $E_{\rm up}/k$ & $\theta_{\rm MB}$ &
$\eta_{\rm MB}$ & Receiver    & PWV        & $T_{\rm{sys}}$\\
&(MHz)    & (K) & (\arcsec)            & & & (mm) & (K)\\
\hline
\CXVIIIO\ $J=3-2$&   329330.553  & 32  &18.0   &  0.75  &  APEX-2a   & $\sim1.0$      & 220  \\
\THCO\ $J=3-2$   &   330587.965  & 32  &18.0   &  0.75  &  APEX-2a   & $\sim1.0$      & 210  \\
\CXVIIIO\ $J=6-5$&   658553.278  & 111 & 9.0   &  0.47  &  \champp\  & $\lesssim0.7$  & 1600  \\
\TWCO\ $J=6-5$   &   691473.076  & 116 & 8.6   &  0.47  &  \champp\  & $\lesssim0.5$  & 1800  \\
\TWCO\ $J=7-6$   &   806651.806  & 155 & 7.4   &  0.45  &  \champp\  & $\lesssim0.5$  & 3900  \\
\THCO\ $J=8-7$   &   881272.808  & 190 & 6.7   &  0.45  &  \champp\  & $\lesssim0.7$  & 3700  \\
 \hline

 \end{tabular}
 
 \tablefoot{Columns are, from left to right, CO isotopologue and transition, frequency, energy
of the transition's upper energy level above the ground-state, the HPBW beam
size, the main beam efficiency, the receiver used, the average precipitable water vapor column and the average system
temperature during the observations. 
\tablefoottext{a}{Taken from the Cologne Database for Molecular Spectroscopy \citep[CDMS,][]{Mueller2005}\footnote{http://www.astro.uni-koeln.de/cdms/}.}
}


 \end{table*}

The Orion Molecular Cloud 1 (OMC-1) is a complex region of the interstellar medium (ISM) stretching over more than 2.4 pc ($20\arcmin$, roughly
north-south) on the sky \citep{Kutner1976}. Its densest part, toward which the Great Orion Nebula (M42), a classical compact ''blister''
\HII\ region, and its associated Orion Nebular Cluster (ONC) appear in projection, is one of the best-studied regions in astronomy. It is a
test bed for studies of (proto)stars and clusters and the formation of low-, intermediate-, and high-mass stars \citep[for a review of this region, see][]{O'Dell2001,O'Dell2008}. Much of this region's prominence is due to its distance of just $414\pm7$ pc \citep{Menten2007}, which makes the ONC and OMC-1 the closest regions of recent (few million years old) and ongoing high-mass star formation. In the following, we shall use the term ``OMC-1 core'' or even just OMC-1 for the roughly $8\arcmin\times8\arcmin$- or 1 pc$^{2}$-sized dense molecular cloud region, which is located closely (0.1--0.2 pc) behind M42 and most of the stars in the ONC \citep[see][the latter give a comprehensive overview of this region and its phenomena]{Zuckerman1973,Genzel1989}.

The OMC-1 core region may be divided into three main zones, all of which show bright (sub)millimeter wavelength emission from warm dust and molecular gas: the Becklin-Neugebauer/Kleinmann-Low (BN/KL) region, Orion South (OMC-1S or Orion-S) and the Orion Bar.

The Orion BN/KL and Orion South regions are considered to be ``twin'' high-mass star forming regions because they have similar masses ($\lesssim$ 100 \Msol) and bolometric luminosities \citep[$10^4-10^5$ \Lsol,][]{Mezger1990,Drapatz1983} and show comparable levels of star-forming activity \citep{O'Dell2008}. The BN/KL region harbors the eponymous ``hot core'' \citep{Masson1984}, which was (and commonly still is) taken to be the prototype of the hot dense regions observed around many newly formed stars \citep[see, e.g.,][]{Kurtz2000}. An interesting alternative explanation for this region's energetics (other than being powered by an embedded central heating source) is a protostellar merger event that released a few times $10^{47}$ erg of energy about 500 years ago \citep{Bally2005,Zapata2011,Bally2011}.

The Orion Bar is a well-described photon-dominated region (PDR) located at the side of the OMC-1 core, facing M42, which is heated and partially ionized by far-ultraviolet (FUV) photons from the young massive stars (most of them from $\theta^1$ C, a spectral type O5--O7 star) that form the ``Trapezium'' at the center of the ONC \citep[see, e.g.,][]{Hollenbach1997,Walmsley2000}. In addition, the Orion Bar appears to be located at the edge of the \HII\ blister tangential to the line of sight.

On giant molecular cloud (GMC) scales, low-rotational level ($J$) \TWCO\ emission (commonly from the $J=1-0$ line) of the ambient, low density gas is usually used to trace the mass of the molecular ISM under a range of assumptions \citep{Bloemen1984,Dame1985}. However, in massive star forming regions with much higher densities and temperatures, observations of the submillimeter and far-infrared (FIR) wavelength mid- or high-$J$ transitions of \TWCO\ and its $^{13}$C and $^{18}$O isotopologues are required for determinations of the gas temperature and density, usually in conjunction with Large Velocity Gradient (LVG) radiative transfer and PDR modelings.

Much of the \TWCO\ emission from the OMC-1 core has been proposed to arise from the neutral and partially ionized back side of the \HII\ blister, i.e., the PDR \citep{Genzel1989}. The pioneering \textit{submillimeter} observations of the \TWCO\ $J=7-6$, $6-5$ and \THCO\ $J=7-6$ transitions by
\citet{Schmid-Burgk1989} and \citet{Graf1990} revealed high density gas ($n\geq$ $10^{4}$ cm$^{-3}$) and elevated temperatures ($T\geq$ 50 K) in all parts of the OMC-1 core region. Observations of the even more highly excited \TWCO\ $J=9-8$ line by \citet{Marrone2004} with the Receiver Lab Telescope (84\arcsec\ resolution) showed that the hot and broad velocity emission in the line profiles arises mainly from the BN/KL region, while much of the narrower ($\sim3$--6 \kms\ wide) emission arises from the PDR. This was also observed in the spectra of the \TWCO\ $J=9-8$ transition taken by \citet{Kawamura2002}, which indicated warm molecular gas from the extended ``quiescent ridge'' region, north of Orion BN/KL. Furthermore, the observations of \TWCO\ $J=7-6$ and $J=4-3$ made by \citet{Wilson2001} together with LVG modeling suggest that the broad line widths of these lines from the BN/KL region are probably due to shock heating, while most of the narrow line extended emission can be explained by PDR models. Recently, \citet{Furuya2009} presented the BN/KL region images in the \TWCO\ $J=7-6$ line.


Clearly, multi-transition, high angular resolution \textit{imaging} studies are needed to disentangle the relative contributions of shock and radiative excitation to the CO emission, and to identify possible energy sources. All of the spectral line observations described above report either single pointing toward selected positions in OMC-1 or limited mapping of special areas such as the Bar \citep{Lis1998} or the BN/KL region \citep{Furuya2009}. The only exceptions are the \TWCO\ $J=7-6$ map by \citet{Schmid-Burgk1989} and the \TWCO\ $J=4-3$ and $J=7-6$ maps by \citet{Wilson2001}. However, the first of these, which covers the whole OMC-1 core, was taken with the poor resolution of the Kuiper Airborne Observatory.

Here we present the first high (better than $10\arcsec$) resolution large scale maps of the OMC-1 core using the $2\times7$ pixel submillimeter \champp\ array receiver on the APEX telescope. Seven of the receiver units cover the 620 to 720 GHz frequency range and seven the 780 to 950 GHz range, allowing imaging of two lines simultaneously. As we shall see in \S\ref{obs}, we imaged the OMC-1 core in \TWCO\ $J=6-5$ plus $J=7-6$ in one run and \THCO\ $J=8-7$ plus \CXVIIIO\ $J=6-5$ in another. The fast mapping afforded by the 14 unit instrument resulted in a superior image consistency. The mid-$J$ CO data were complemented by the maps of the \THCO\ and \CXVIIIO\ $J=3-2$ lines.


\section{\label{obs}Observations}

 \begin{figure*}
 \centering
 \includegraphics[angle=-90,width=0.95\textwidth]{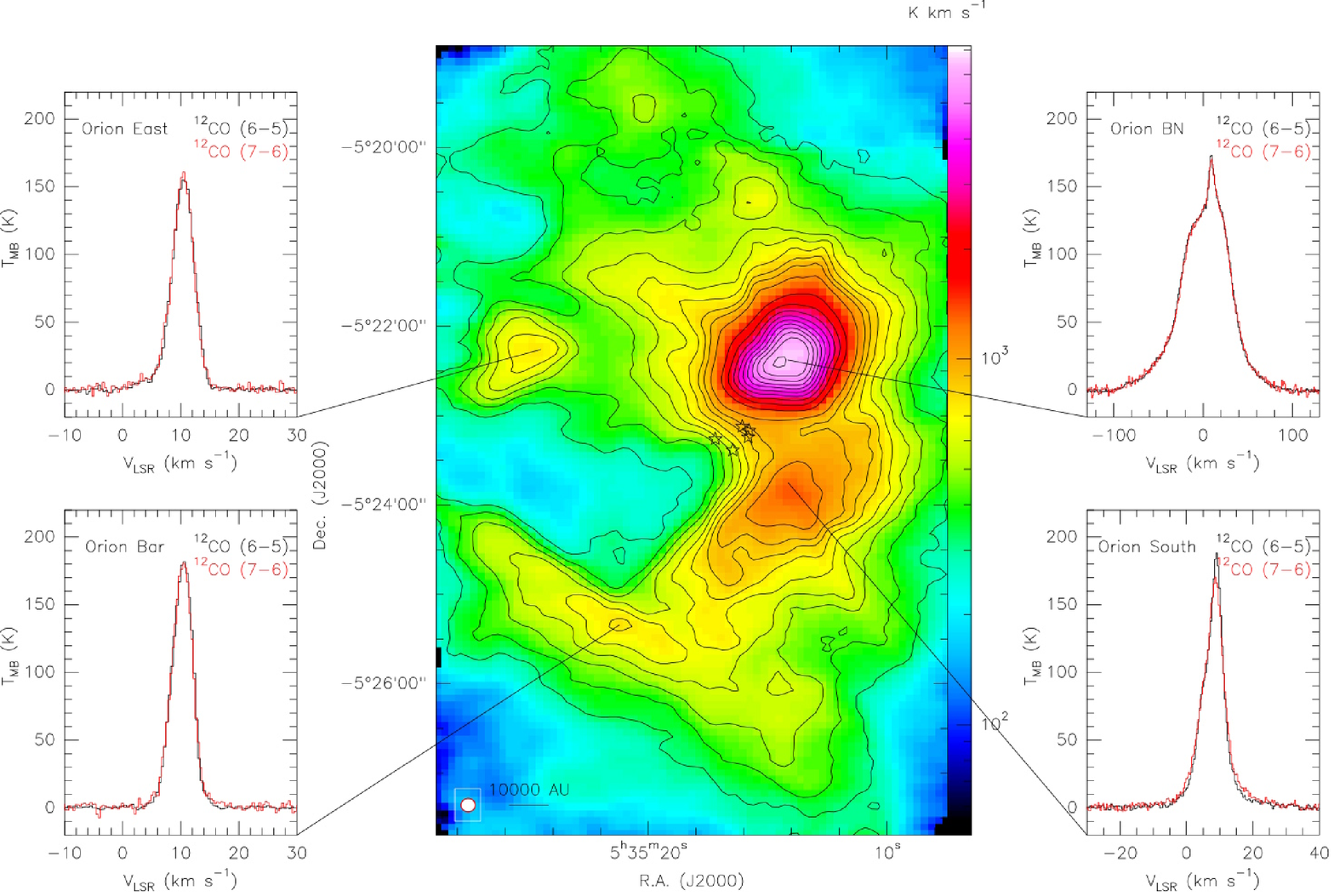}
 \caption{The OMC-1 core \TWCO\ $J=6-5$ integrated intensity [$-25$, $+30$] \kms\ image overlaid with the $J=7-6$ integrated intensity [$-25$, $+30$] \kms\ contours running from 300 to 1000 K \kms\ in steps of 100 K \kms, and the subsequent contours are plotted from 1200 to 7200 K \kms\ in steps of 600 K \kms. The \TWCO\ $J=6-5$ (black) and \TWCO\ $J=7-6$ (red) spectra are shown for Orion BN, Orion South, Orion Bar, and Orion East. Both images have been smoothed using Gaussian profiles (a width of 9\farcs1 for \TWCO\ $J=6-5$ and 7\farcs8 for \TWCO\ $J=7-6$) from their original Nyquist-sampled images. The black stars mark the positions of the five Trapezium stars ($\theta^1$ Ori A, B, C, D, and E). The original beam sizes are shown at the bottom left of the image for \TWCO\ $J=6-5$ and $J=7-6$ in red (9\farcs6) and white (8\farcs2), respectively.\label{map1}}
 \end{figure*}

The observations were performed in 2007 November and 2008 July with the 12 meter APEX telescope on Llano de Chajnantor in Chile \citep{Gusten2006}\footnote{This publication is based on the data acquired with the Atacama Pathfinder Experiment (APEX) telescope. APEX is a collaboration between the Max-Planck-Institut f\"ur Radioastronomie, the European Southern Observatory, and the Onsala Space Observatory.}.

Large-scale maps of the \THCO\ and \CXVIIIO\ $J=3-2$ lines were obtained in the On-The-Fly (OTF) observations with the APEX-2a facility receiver \citep{Risacher2006}. This double sideband (DSB) heterodyne receiver provided receiver DSB temperatures of 60 K and typical system noise from 100 to 200 K. Maps of size $\sim300\arcsec\times350\arcsec$ were centered on the Orion BN source ($\alpha,\delta_{J2000}= 05^{\rm{h}}35^{\rm{m}}14\fs16$, $-05\degr22\arcmin21\farcs5$). For all spectra, position switching to an emission-free reference position at offsets ($-500$\arcsec, 0\arcsec) was used. The Fast Fourier Transform Spectrometers \citep[FFTS,][]{Klein2006} were operated with a 1000 MHz bandwidth and a $\sim$ 0.1 km s$^{-1}$ spectral resolution.


The submm observations of the high-$J$ CO lines were performed with the dual-color heterodyne array \champp\ \citep{Kasemann2006,Gusten2008}, operating $2\times7$ elements simultaneously in the 450 \micron\ and 350 \micron\ atmospheric windows. We observed two setups: in 2007 November the \TWCO\ $J=6-5$ and $J=7-6$ lines were mapped simultaneously, while in the second coverage in 2008 July, the \THCO\ $J=8-7$ and \CXVIIIO\ $J=6-5$ transitions were observed in parallel. The MPIfR Array Correlator Spectrometer (MACS) was used as the backend during the former run, and the central pixels of each array were always connected to the FFT spectrometers mentioned above. Resolutions were $\sim$ 0.42 km s$^{-1}$ for MACS and $\sim$ 0.05 km s$^{-1}$ for the FFTS. For the 2008 July observations, MACS had been replaced with a new FFTS array backend, operating for each pixel with a $2\times1.5$ GHz bandwidth and 8192 channels per FFTS. We mapped the OMC-1 core region in a size of $6\arcmin\times8\arcmin$ ($0.72\times0.96$ pc$^2$) in OTF slews in the right ascension direction (each $\sim3\arcmin$ long), dumping data every $4\arcsec$. Subsequent scans were also spaced by $4\arcsec$ in declination so that the Nyquist sampled maps for each pixel were obtained. The angular resolution varies with frequency between $6\farcs7$ (881 GHz) and $9\farcs0$ (658 GHz). The same reference position at offsets ($-500\arcsec, 0\arcsec$) was used. The on-source integration time per dump and per pixel was 1 second only. With the above over-sampling strategy, all pixels of \champp\ covered a given grid position at least once.

 \begin{figure*}
 \centering
 \includegraphics[angle=270,width=0.95\textwidth]{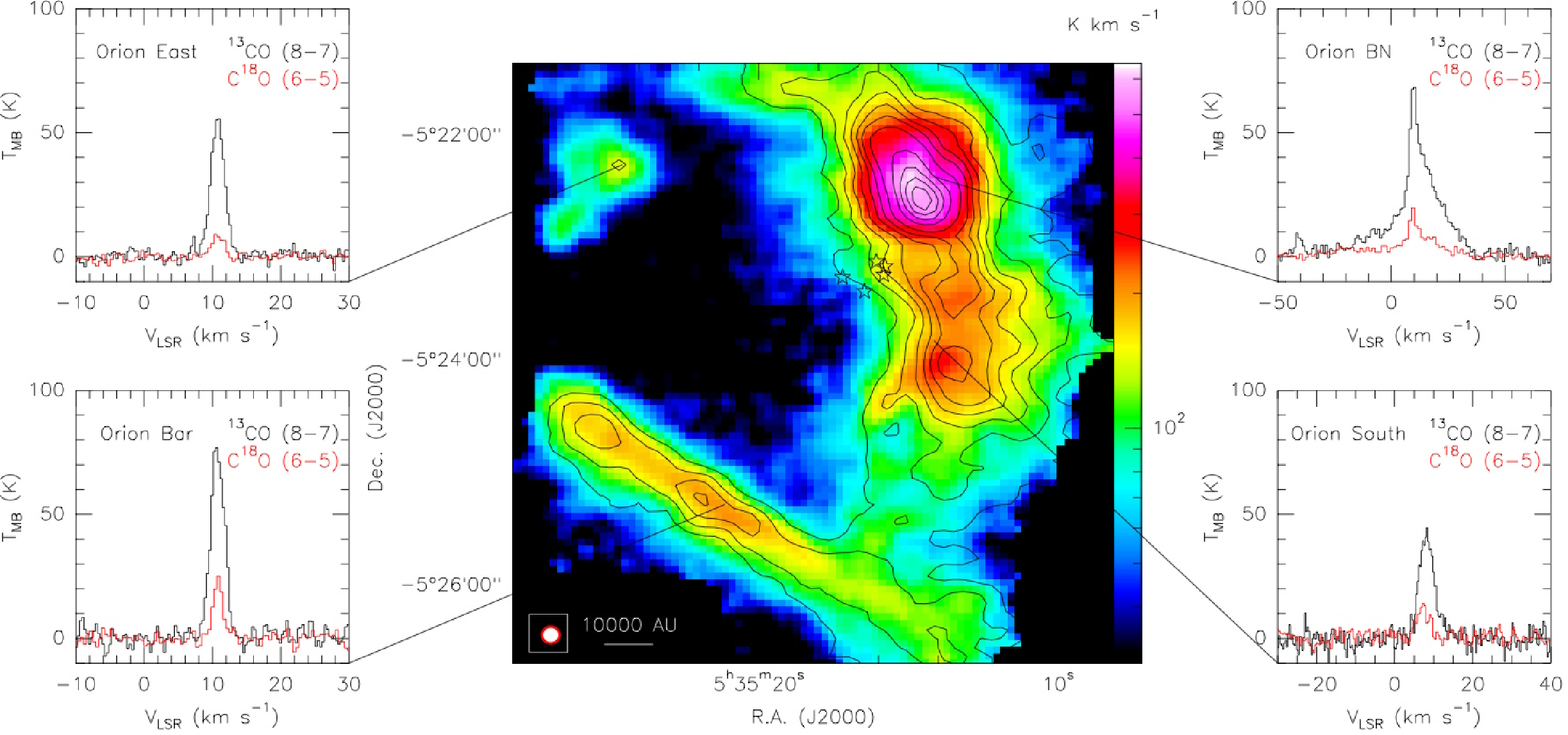}
 \caption{The OMC-1 core \THCO\ $J=8-7$ integrated intensity [$+$5, $+$15] \kms\ image overlaid with the \CXVIIIO\ $J=6-5$ [$+$4, $+$12] \kms\ contours running from 15 to 65 K \kms\ in steps of 10 K \kms, and the subsequent contours are plotted from 80 to 200 K \kms\ in steps of 30 K \kms. The \THCO\ $J=8-7$ (black) and \CXVIIIO\ $J=6-5$ (red) spectra are shown for Orion BN, Orion South, Orion Bar, and Orion East at the same positions as in Figure \ref{map1}. Both images have been smoothed using Gaussian profiles (a width of 7\farcs1 for \THCO\ $J=8-7$ and 9\farcs5 for \CXVIIIO\ $J=6-5$) from their original Nyquist-sampled images. The black stars mark the positions of the five Trapezium stars. The original beam sizes are shown at the bottom left of the image for \THCO\ $J=8-7$ and \CXVIIIO\ $J=6-5$ in red (7\farcs5) and white (10\farcs1), respectively.\label{map2}}
 \end{figure*}

Data calibration was performed every $\sim$10 minutes with cold and ambient temperature loads. The data were processed with the APEX online calibrator, assuming an image sideband suppression of 10 dB. The \champp\ observing parameters, the telescope beam sizes, and typical values for the Precipitable Water Vapor (PWV) are summarized in Table \ref{table1}. Typical single sideband (SSB) system temperatures were well below 2000 K and 4000 K for the low- and high-frequency array of \champp, respectively.

Telescope pointing was established by \THCO\ and \CXVIIIO\ $J=3-2$ line pointings every hour on nearby R Lup for the APEX-2a observations, resulting in an accuracy of $\sim2\arcsec$. In the shorter-wavelength submm windows, pointing is more difficult because of the much reduced system sensitivities and the paucity of suitable pointing sources, strong enough for peak-up on their continuum and/or line emission. During the 2007 \champp\ observations, $o$ Cet was used as a line pointing source in the \TWCO\ $J=6-5$ mapping, but with an on-sky distance of $\approx50^\circ$ to the OMC-1 core, the absolute pointing on the latter was not accurately corrected. The wider IF processors and backends that became available in 2008 improved the situation significantly: simultaneous observations of the \TWCO\ $J=6-5$ line and the CH$_3$CN $J_k = 36_9-35_9$ multiplet (near 660.6 GHz) became possible in the low-frequency array of \champp. We assume that the highly rotationally excited (lower energy level $E/k$=133 K above the ground state) methyl cyanide line emission does peak on the Orion Hot Core proper, as defined by the centroid of 92 GHz CH$_3$CN $J_k=5_k-4_k$ emission determined with the BIMA interferometer, i.e., $\alpha,\delta_{\rm J2000} = 05^{\rm{h}}35^{\rm{m}}14\fs48$, $-05\degr22\arcmin30\farcs6$ \citep{Wilner1994}. Thereby, we established a pointing reference local to the OMC-1 core complex. While those small pointing maps toward the Orion Hot Core were made, the \TWCO\ $J=7-6$ transition in the high-frequency array was observed in parallel, and provided the link to the 2007 observations of the main CO isotopologues. The central pixels between the two arrays are co-aligned (better than one arcsecond). In practice, we shifted the 2007 maps such that the \TWCO\ $J=7-6$ line wing emission from $-50$ to $-30$ km~s$^{-1}$ match the pointed 2008 observations. The overall accuracy in the higher-$J$ CO observations with \champp\ is estimated to be $\lesssim4\arcsec$.

\begin{figure*}
 \centering
 \includegraphics[angle=270,width=\textwidth]{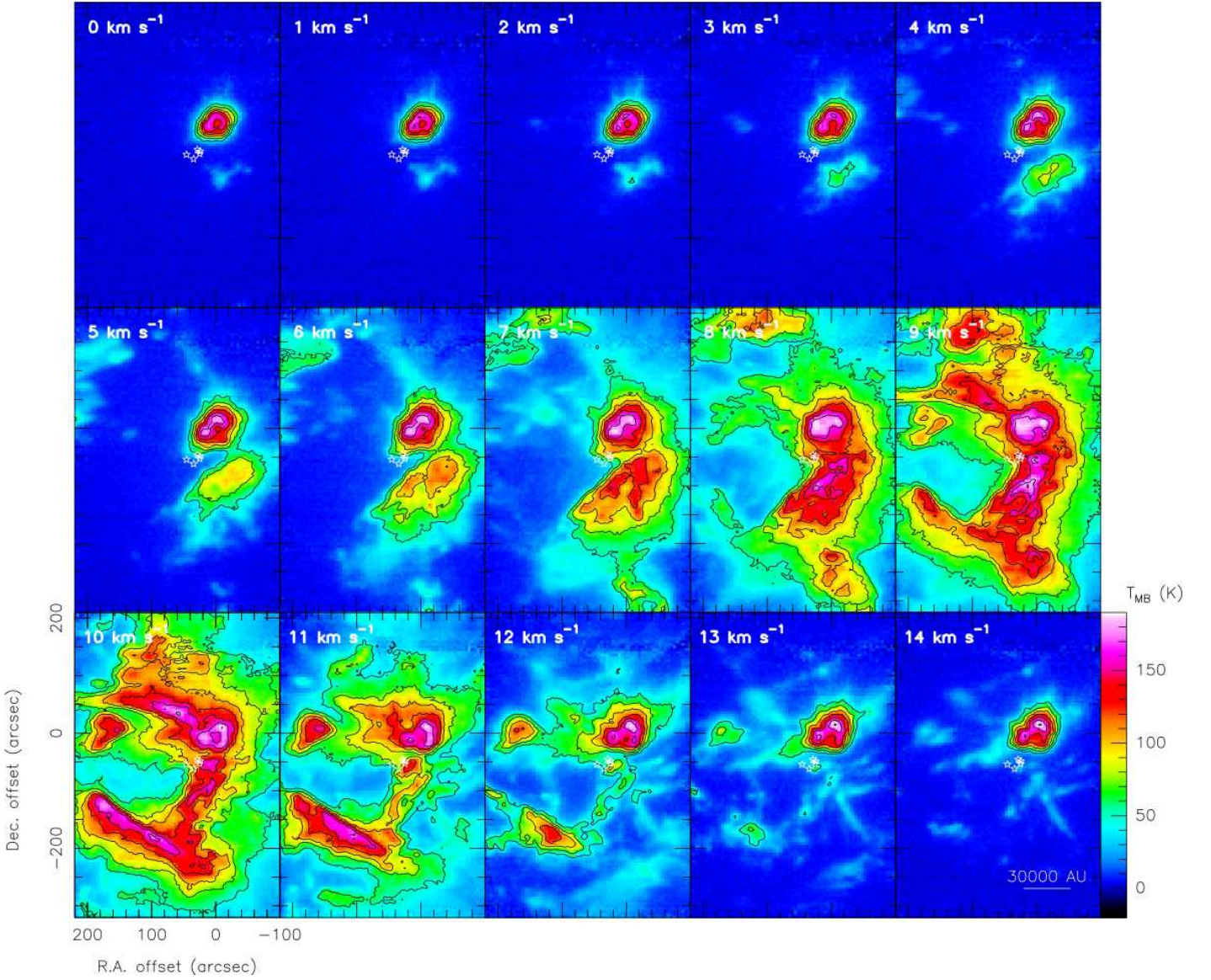}
 \caption{Low-velocity channel maps of the \TWCO\ $J=6-5$ line with the $J=7-6$ contours in the OMC-1 core. The contours are running from 50 to 200 K in steps of 25 K. Orion BN is located at the offset (0\arcsec, 0\arcsec), and white stars mark the positions of the five Trapezium stars. \label{chmap}}
 \end{figure*}

All spectra were converted to the main beam brightness temperature unit, $T_{\rm MB}=T_{\rm A}^{*}/\eta_{\rm MB}$ ($\eta_{\rm MB}=B_{\rm eff}/F_{\rm eff}$), using a forward efficiency ($F_{\rm eff}$) of 0.95 and beam coupling efficiencies ($B_{\rm eff}$) as determined toward Jupiter (Table \ref{table1}). The latter is motivated by the average size of the emission in velocity space (Fig. \ref{chmap}). All data were reduced using the standard procedures in the CLASS and GREG programs from the GILDAS package\footnote{http://www.iram.fr/IRAMFR/GILDAS/}.

 \begin{figure*}
 \centering
 \includegraphics[angle=270,width=0.98\textwidth]{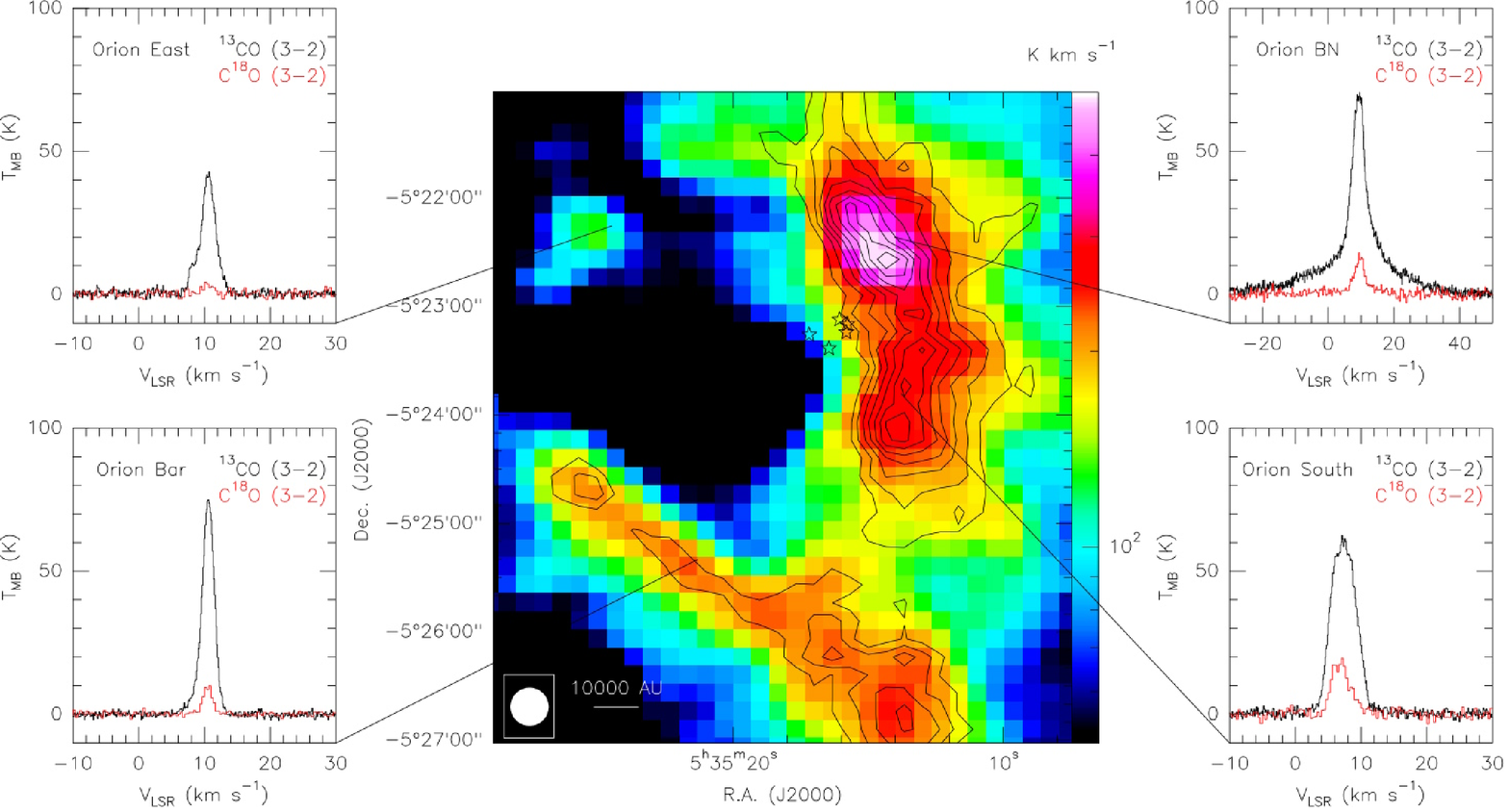}
 \caption{The OMC-1 core \THCO\ $J=3-2$ integrated intensity [$+$5, $+$15] \kms\ image overlaid with the \CXVIIIO\ $J=3-2$ [$+$4, $+$12] \kms\ contours running from 24 to 80 K \kms\ in steps of 8 K \kms. The \THCO\ $J=3-2$ (black) and \CXVIIIO\ $J=3-2$ (red) spectra are shown for Orion BN, Orion South, Orion Bar, and Orion East at the same positions as in Fig. \ref{map1}. The black stars mark the positions of the five Trapezium stars. The beam size of 21\farcs1 is shown at the bottom left of the image.\label{map3}}
 \end{figure*}

\begin{table*}
\caption{CO isotopologues measurements in the selected sources of the OMC-1 core}             
\label{table2}      
\centering
\renewcommand{\footnoterule}{}
\begin{tabular}{lcccccc}     
\hline\hline       
Molecule/Line & \multicolumn{3}{c}{Orion BN} & \multicolumn{3}{c}{Orion South} \\ 
\hline
& $T_{\rm peak}$ & $V_{\rm LSR}$ & $\Delta V$ & $T_{\rm peak}$ & $V_{\rm LSR}$ & $\Delta V$  \\
& (K)            & (\kms)        & (\kms)     & (K)            & (\kms)        & (\kms)    \\
\hline                    

\THCO\ $J=8-7$    & 63.8$\pm$2.3  & 10.3$\pm$0.3 & 9.0$\pm$0.3  & 41.0$\pm$3.6  & 8.1$\pm$0.3 & 4.9$\pm$0.3  \\ 
\TWCO\ $J=7-6$    & 168.5$\pm$2.2 & 9.8$\pm$1.1  & 55.5$\pm$1.1 & 165.6$\pm$2.0 & 8.7$\pm$1.1 & 6.9$\pm$1.1  \\ 
\TWCO\ $J=6-5$    & 164.6$\pm$1.6 & 9.5$\pm$1.3  & 53.3$\pm$1.3 & 187.5$\pm$1.4 & 9.1$\pm$1.3 & 6.4$\pm$1.3  \\ 
\CXVIIIO\ $J=6-5$ & 19.0$\pm$1.3  & 10.0$\pm$0.4 & 5.8$\pm$0.4  & 10.2$\pm$1.4  & 7.6$\pm$0.4 & 4.4$\pm$0.4  \\ 
\THCO\ $J=3-2$    & 68.7$\pm$0.5  & 9.6$\pm$0.3  & 5.6$\pm$0.3  & 62.1$\pm$1.1  & 7.3$\pm$0.3 & 4.5$\pm$0.3  \\ 
\CXVIIIO\ $J=3-2$ & 12.4$\pm$0.7  & 10.2$\pm$0.3 & 3.0$\pm$0.3  & 19.8$\pm$1.6  & 7.0$\pm$0.3 & 3.3$\pm$0.3  \\
\hline                  

\hline
 &  \multicolumn{3}{c}{Orion Bar} & \multicolumn{3}{c}{Orion East}\\ 
\hline

\THCO\ $J=8-7$    &  75.9$\pm$3.3  & 10.7$\pm$0.3 & 2.7$\pm$0.3  & 55.7$\pm$2.5   &  10.6$\pm$0.3  &  2.3$\pm$0.3 \\ 
\TWCO\ $J=7-6$    &  179.9$\pm$2.6 & 10.5$\pm$1.1 & 4.0$\pm$1.1  & 156.2$\pm$3.1  &  10.5$\pm$1.1  &  4.0$\pm$1.1 \\ 
\TWCO\ $J=6-5$    &  180.9$\pm$0.9 & 10.4$\pm$1.3 & 3.8$\pm$1.3  & 156.7$\pm$1.2  &  10.4$\pm$1.3  &  4.7$\pm$1.3 \\ 
\CXVIIIO\ $J=6-5$ &  23.9$\pm$2.2  & 10.9$\pm$0.4 & 1.8$\pm$0.4  &  8.2$\pm$1.3   &  10.4$\pm$0.4  &  2.2$\pm$0.4 \\ 
\THCO\ $J=3-2$    &  90.1$\pm$1.1  & 10.8$\pm$0.3 & 2.4$\pm$0.3  & 41.6$\pm$1.1   &  10.6$\pm$0.3  &  2.3$\pm$0.3 \\ 
\CXVIIIO\ $J=3-2$ &  13.4$\pm$1.0  & 10.8$\pm$0.3 & 1.7$\pm$0.3  &  3.9$\pm$1.0   &  10.4$\pm$0.3  &  2.3$\pm$0.3 \\
\hline
\end{tabular}

\tablefoot{Temperatures shown here are corrected to the main beam temperature unit, and line widths are measured in FWHM.
}

%

\end{table*}

 \begin{figure*}
 \centering
  \includegraphics[angle=270,width=\textwidth]{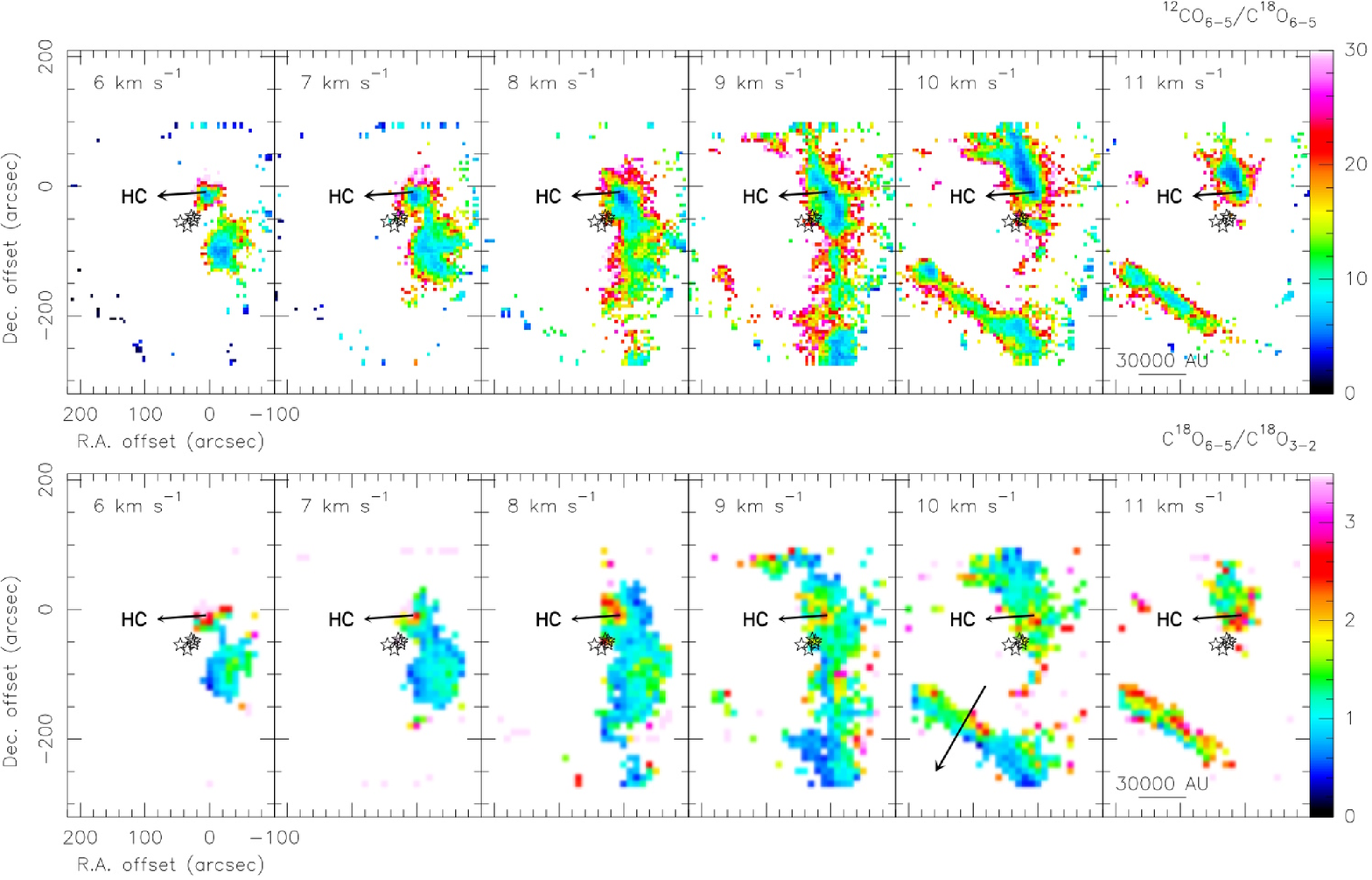}
  \caption{Upper panel is the \TWCO\ $J=6-5$ to \CXVIIIO\ $J=6-5$ ratio channel maps ($\theta_{\rm HPBW}=10\arcsec$) at different velocities from 6 to 11 \kms\ in the OMC-1 core. The edge cut and empty area are due to different image sizes and data blanking (2 $\sigma$ level of the \CXVIIIO\ $J=6-5$ temperature). Lower panel is the corresponding \CXVIIIO\ $J=6-5$ to $J=3-2$ ratio channel maps ($\theta_{\rm HPBW}=20\arcsec$). The black arrow in the 10 \kms\ image indicates the ratio gradient across the Orion Bar (see text). The Orion Hot Core (HC) position is marked, and Orion BN is located at the offset (0\arcsec, 0\arcsec). The positions of the five Trapezium stars are marked in each map.\label{ratio3}}
 \end{figure*}


\section{Results and discussion}

\subsection{Large scale maps -- overall morphology}

Figure \ref{map1} shows $\approx6\arcmin\times8\arcmin$-size maps of the OMC-1 core in the intensity of the \TWCO\ $J=6-5$ and $J=7-6$
lines integrated over the local standard of rest (LSR) velocity interval from $-$25 to $+$30~\kms. The emission distributions of both lines, clearly resolved in our observations, resemble each other closely and are consistent with the JCMT/SCUBA 850 \micron\ image obtained by \citet{Johnstone1999}, where several regions are discernible. The most prominent of which are: (a) the emission maximum centered on the BN/KL region, (b) a secondary maximum coinciding with Orion South (OMC-1S), (c) a region $\approx3\arcmin$ east of the Trapezium, which is known as Orion East (OMC-1 E; see below), and (d), the prominent straight Orion Bar PDR. The bar actually appears connected to the Orion South region, forming a mirrored letter L shape. Furthermore, there are two minor maxima at $1\farcm8$ N and $3\farcm3$ NNE of the BN/KL peak position.

Apart from their spatial distributions, the regions above can be distinguished by their centroid LSR velocities, $V_{\rm LSR}$, and their line widths, $\Delta V$. The Orion BN/KL region displays a spectacularly broad line up to $200$~\kms\ wide, centered around 6~\kms\ with a narrow ($\Delta V \approx 5$~\kms) feature at $V_{\rm LSR}\approx8$~\kms\ superposed. Orion South is characterized by $\sim6$~\kms\ wide lines, with higher values at positions with outflow activities. The \TWCO\ $J=6-5$ and $J=7-6$ lines observed toward the Orion Bar have FWHM widths of $\sim4$~\kms\ and are centered around $V_{\rm LSR}=10$~\kms. The Orion East region has previously been mapped by \citet{Houde2004} and \citet{Herrmann1997}, but otherwise has received relatively little attention. The observed line widths and LSR velocities of Orion East, as well the high intensity of its CO emission ($T_{\rm{MB}}\gtrsim 160$ K) clearly demonstrate its PDR nature. All our measurements are in good agreement with the results of \citet{Schmid-Burgk1989}, \citet{Graf1990}, \citet{Wilson2001}, \citet{Kawamura2002}, and \citet{Marrone2004}. The measurements of the CO isotopologues in the OMC-1 core are summarized in Table \ref{table2}.


Figure \ref{map2} shows the integrated intensity of \THCO\ $J=8-7$ and \CXVIIIO\ $J=6-5$ maps in a similar area as \TWCO\ but in a smaller range of radial velocities (from $+$5 to $+$15 km s$^{-1}$ for \THCO\ and from $+$4 to $+$12 km s$^{-1}$ for \CXVIIIO). Both maps also show a good correspondence but with the integrated emission from \CXVIIIO\ $J=6-5$ fainter than \THCO\ $J=8-7$. The two isotopologues have fainter and less broader lines compared with the \TWCO\ line emission which is expected for optically thick lines (see the spectra in Figs. \ref{map1} and \ref{map2}). The Orion Bar is well resolved in both lines and also shows a sharp edge. In the mid-$J$ images (Figs. \ref{map1} and \ref{map2}), the strong peak in the center of the Bar is also seen in the recent {\it Herschel}/SPIRE high-$J$ \TWCO\ and \THCO\ data obtained by \citet{Habart2010}. In addition, the line widths of \THCO\ $J=8-7$ and \CXVIIIO\ $J=6-5$ toward all four main regions are narrow, even in the outflow zones, i.e., Orion BN/KL and South.

The \THCO\ $J=3-2$ and \CXVIIIO\ $J=3-2$ integrated intensity maps are shown in Figure \ref{map3}. These images were made in a similar area as higher-$J$ \TWCO\ and in a range of radial velocities equal to \THCO\ $J=8-7$ and \CXVIIIO\ $J=6-5$. The \CXVIIIO\ $J=3-2$ emission is only present where the \THCO\ $J=3-2$ emission is strong. The \THCO\ and \CXVIIIO\ line profiles at this rotational transition toward the Orion BN/KL and South regions are very broad and show well-defined wings. The southwestern part of the Bar is especially strong in this lower-$J$ transition for both isotopologues, which is also shown in the {\it Herschel}/SPIRE high-$J$ \TWCO\ and \THCO\ images reported by \citet{Habart2010}.

\subsection{Velocity structure}

The \TWCO\ $J=6-5$ and $J=7-6$ velocity channel maps are shown in Figure \ref{chmap}. These maps reveal a well-resolved and complex structure of the OMC-1 core at ambient velocities as well as higher radial velocities. The outflows that are located mainly at Orion BN/KL, Orion South, and toward the north of Orion BN/KL are clearly seen in these images \citep[see][]{Zapata2006,Zapata2009}, e.g., \vlsr=0--6 \kms\ for the BN/KL region and \vlsr=12--14 \kms\ for the Orion South region. Besides, no clear north-south velocity gradients across the OMC-1 core region are observed.


The compact and warm structures toward the BN/KL region traced by the \TWCO\ $J=6-5$ and $J=7-6$ lines, which display a very broad range of velocities (up to about $\pm100$ \kms), are part of the enigmatic molecular outflow that seems to be produced by a violent explosion during the disruption of a massive young stellar system \citep{Bally2005,Zapata2009,Bally2011}. Our observations reveal that some faint filamentary structures are very likely associated with the high-velocity \TWCO\ bullets (Peng et al, in prep.) reported by \citet{Zapata2009}. 



\subsection{Line ratios}

\begin{figure}
 \centering
 \includegraphics[angle=270,width=0.48\textwidth]{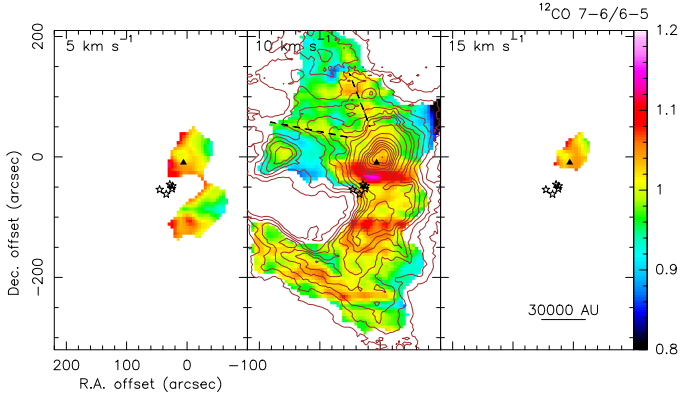}
 \caption{Velocity channel maps (5, 10, and 15 km s$^{-1}$) of the \TWCO\ $J=7-6$ to $J=6-5$ ratio ($\theta_{\rm HPBW}=10\arcsec$) in the OMC-1 core. The 10 \kms\ image is overlaid with the \TWCO\ $J=6-5$ contours as shown in Fig. \ref{map1}. The empty area are due to data blanking (3 $\sigma$ level of the \TWCO\ $J=7-6$ temperature). The dashed-lines indicate that these ratio gradients are likely caused by outflows or filaments. Black triangles mark the Orion Hot Core position. \label{ratio1}}
 \end{figure}

\begin{figure}
 \centering
 \includegraphics[angle=270,width=0.45\textwidth]{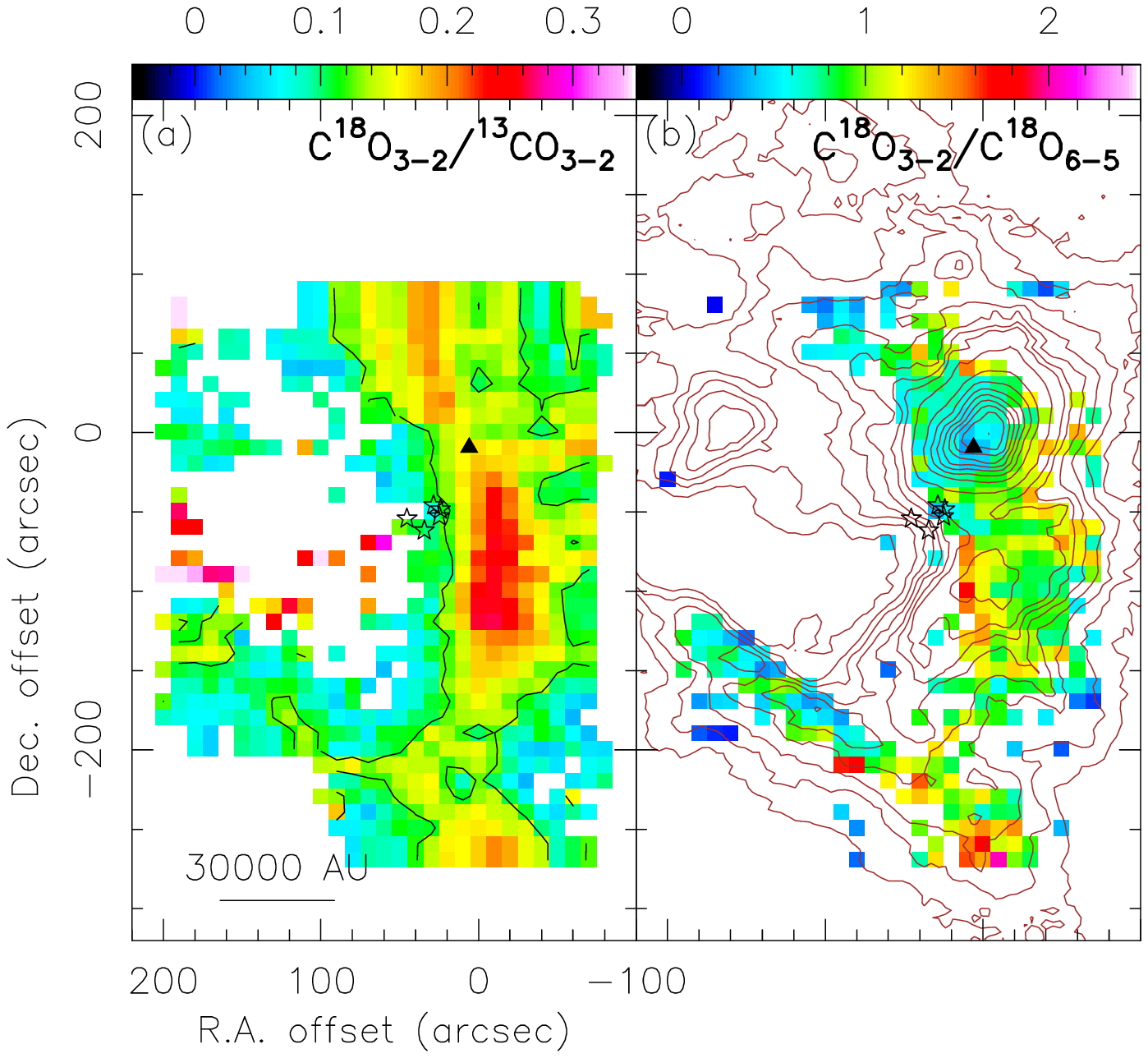}
 \caption{(a) The OMC-1 core \CXVIIIO\ $J=3-2$ to \THCO\ $J=3-2$ integrated intensity ratio map ($\theta_{\rm HPBW}=20\arcsec$) overlaid with the \TWCO\ $J=6-5$ contours (grey) as shown in Fig. \ref{map1}. The black contour represents the \CXVIIIO/\THCO\ abundance ratio of 0.122 (60/490). (b) The OMC-1 core \CXVIIIO\ $J=3-2$ to $J=6-5$ integrated intensity ratio map ($\theta_{\rm HPBW}=20\arcsec$) overlaid with the \TWCO\ $J=6-5$ contours as shown in Fig. \ref{map1}. Black triangles mark the Orion Hot Core position. \label{ratio2}}
 \end{figure}

In Figures \ref{ratio3}--\ref{ratio2}, we show the line intensity ratio maps between the CO isotopologues and the different rotational transitions.

From the image of the \TWCO\ to \CXVIIIO\ $J=6-5$ ratio (Fig. \ref{ratio3} upper panel), we can see the spatial distribution of high optical depth regions at different velocities. The north-south dense ridge from Orion BN/KL to Orion South is clearly seen at 9 \kms. This high optical depth ridge is similar to the filamentary structures of the NH$_{3}$ emission \citep{Wiseman1996,Wiseman1998}, but has some morphological differences in the north of the Orion BN/KL region. Besides, the straight shape of the Orion Bar is pronounced at \vlsr=10--11 \kms, where three high optical depth regions are seen at the two ends and center of the Bar at \vlsr=10 and 11 \kms, respectively. In the lower panel of Figure \ref{ratio3}, the \CXVIIIO\ $J=6-5$ to $J=3-2$ ratios around 10--11 \kms\ show a gradient in the Orion Bar which goes in the direction from the Trapezium stars. It indicates that \CXVIIIO\ $J=6-5$ is strongly excited at the edge of the Bar by UV photons from the Trapezium stars.


Figure \ref{ratio1} shows the ratio between \TWCO\ $J=7-6$ and $J=6-5$ at three different radial velocities of 5, 10, and 15 \kms. The three panels show clear variations across the integrated line emission. It is interesting to note that at the cloud velocity of 10 km s$^{-1}$, there are high 7--6/6--5 ratios located very close to the position of the Trapezium stars toward Orion BN/KL. These gradients are likely produced by those massive stars that heat the molecular cloud, causing the stronger \TWCO\ $J=7-6$ line emission compared with the $J=6-5$ line. A gradient is also seen in the Orion Bar which goes in the perpendicular direction to the Trapezium stars (Fig. \ref{ratio1} at $V_{\rm LSR}=10$ \kms), and some patches with higher \TWCO\ $J=7-6$ brightness temperatures are seen inside the Bar or behind the ionization front. Some horizontal and vertical strips, artifacts from the OTF mapping, are also seen in the Orion Bar and South regions. These artifacts can affect the intensity ratio by $\sim 14 \%$ given the calibration errors of $10 \%$ for both \TWCO\ $J=6-5$ and $J=7-6$. However, some structures seen in the \TWCO\ $J=7-6$ to $J=6-5$ ratio maps, e.g., the dashed-lines in Figure \ref{ratio1}, are unlikely due to the calibration error or scanning strips during the observation, and seem to be footprints of filaments or outflows. Some similar filament structures have been noticed in the dust continuum emission observed by \citet{Johnstone1999}.


In the ratio map of \CXVIIIO\ $J=3-2$ to \THCO\ $J=3-2$ shown in Figure \ref{ratio2}, a clear elongated and dense ridge is seen in the north-south direction, where the \THCO\ $J=3-2$ intensity is weaker in the Orion BN/KL region than at Orion South and the north of BN/KL. Therefore, the lower CO column density derived from the lower \CXVIIIO\ $J=3-2$ to \THCO\ $J=3-2$ ratio toward Orion BN/KL may be misleading. Even though the \THCO\ $J=3-2$ line is optically thick and may be self-absorbed, this will result in a higher \CXVIIIO\ $J=3-2$ to \THCO\ $J=3-2$ ratio instead of a lower ratio in Orion BN/KL. As \citet{Goldsmith1997} pointed out, the column density calculated using low-$J$ \CXVIIIO\ lines is accurate except for the temperature $\geq150$ K, which is the case in the OMC-1 core. Hence, higher-$J$ CO observations are critical to determine the CO column density in the OMC-1 core region. In addition, already shown in Figure \ref{ratio1}, the low \CXVIIIO\ $J=3-2$ to $J=6-5$ ratios indicate that the \CXVIIIO\ $J=6-5$ intensity is strongly enhanced toward the Orion BN/KL, South, and Bar regions, and is likely due to their PDR nature. Nevertheless, shocks/outflows may also play an important role in the Orion BN/KL and South regions, leading to an extra heating in the cloud.



\subsection{Excitation temperature and density estimates}


The \TWCO\ $J=6-5$ and \CXVIIIO\ $J=6-5$ data are used here to derive the gas excitation temperature and density of the OMC-1 core. In the submillimeter regime where the Rayleigh-Jeans approximation is often not valid, the observed radiation temperature in local thermodynamic equilibrium (LTE) can be expressed as
\begin{equation}\label{cal-orion-1}
{T^{*}_{\rm R}}=\frac{h\nu}{k}\biggl[\frac{1}{e^{h\nu/kT_{\rm ex}}-1}-\frac{1}{e^{h\nu/kT_{\rm bg}}-1}\biggr](1-e^{-\tau_{\nu}}),
\end{equation}
where $T_{\rm ex}$ is the excitation temperature and $\tau_{\nu}$ is the optical depth at a specific molecular line transition. The background brightness temperature $T_{\rm bg}$ includes the cosmic background radiation of 2.73 K and the radiation from warm dust, ranging from about 5 K for less dense gas to about 26 K for the Orion Hor Core \citep{Goldsmith1997}. The second term inside the brackets can be neglected since it only contributes $\lesssim2\%$ of $T^{*}_{\rm R}$. By making the assumption of the same excitation temperature for \TWCO\ and \CXVIIIO\ $J=6-5$, the optical depths of \TWCO\ and \CXVIIIO\ $J=6-5$ can be determined from the relation
\begin{equation}\label{cal-orion-2}
\frac{T^{*}_{\rm R}(\rm{^{12}CO)}}{T^{*}_{\rm R}(\rm{C^{18}O)}}\approx\frac{1-e^{-\tau(\rm{^{12}CO})}}{1-e^{-\tau(\rm{C^{18}O})}}.
\end{equation}
Since the optical depth of \TWCO\ $J=6-5 \gg 1$, the optical depth of \CXVIIIO\ $J=6-5$ can be directly obtained ($\sim 0.08$) by assuming that the optical depth ratio is approximated to the isotopologic abundance ratio, and here we adopted [\TWCO]/[\CXVIIIO] of 490 \citep[][]{Boreiko1996,Wilson1992} and a beam filling factor of unity ($T^{*}_{\rm R}=T_{\rm MB}$). The excitation temperatures of \TWCO\ and \CXVIIIO\ $J=6-5$ can be estimated via
\begin{equation}\label{cal-orion-3}
T_{\rm ex}=\frac{h\nu}{k}\biggl[{\rm ln}\biggl(1+\frac{h\nu}{kT_{\rm MB}\rm(^{12}CO)}\biggr)\biggr]^{-1}.
\end{equation}
Then the total \CXVIIIO\ column density can be derived from
\begin{equation}
N({\rm C^{18}O})=\frac{3kQ_{\rm rot}}{8\pi^{3}\nu S\mu^{2}}e^{\frac{E_{\rm up}}{kT_{\rm ex}}}\int T_{\rm MB}dV
\end{equation}
\begin{equation}
\simeq1.33\times10^{12}\ (T_{\rm ex}+0.88)\ e^{\frac{E_{\rm up}}{kT_{\rm ex}}}\int T_{\rm MB}dV\ \rm{cm^{-2}},
\end{equation}
where $E_{\rm up}$ is 110.6 K for the \CXVIIIO\ $J=6-5$ transition and $Q_{\rm rot}$ is the rotational partition function. In the end, the \HH\ column density can be estimated by adopting a [\TWCO]/[\CXVIIIO] abundance ration of 490 and a \TWCO\ abundance of $8\times10^{-5}$ \citep{Wilson1992}.


  \begin{figure*}
 \centering
 \includegraphics[angle=270,width=0.9\textwidth]{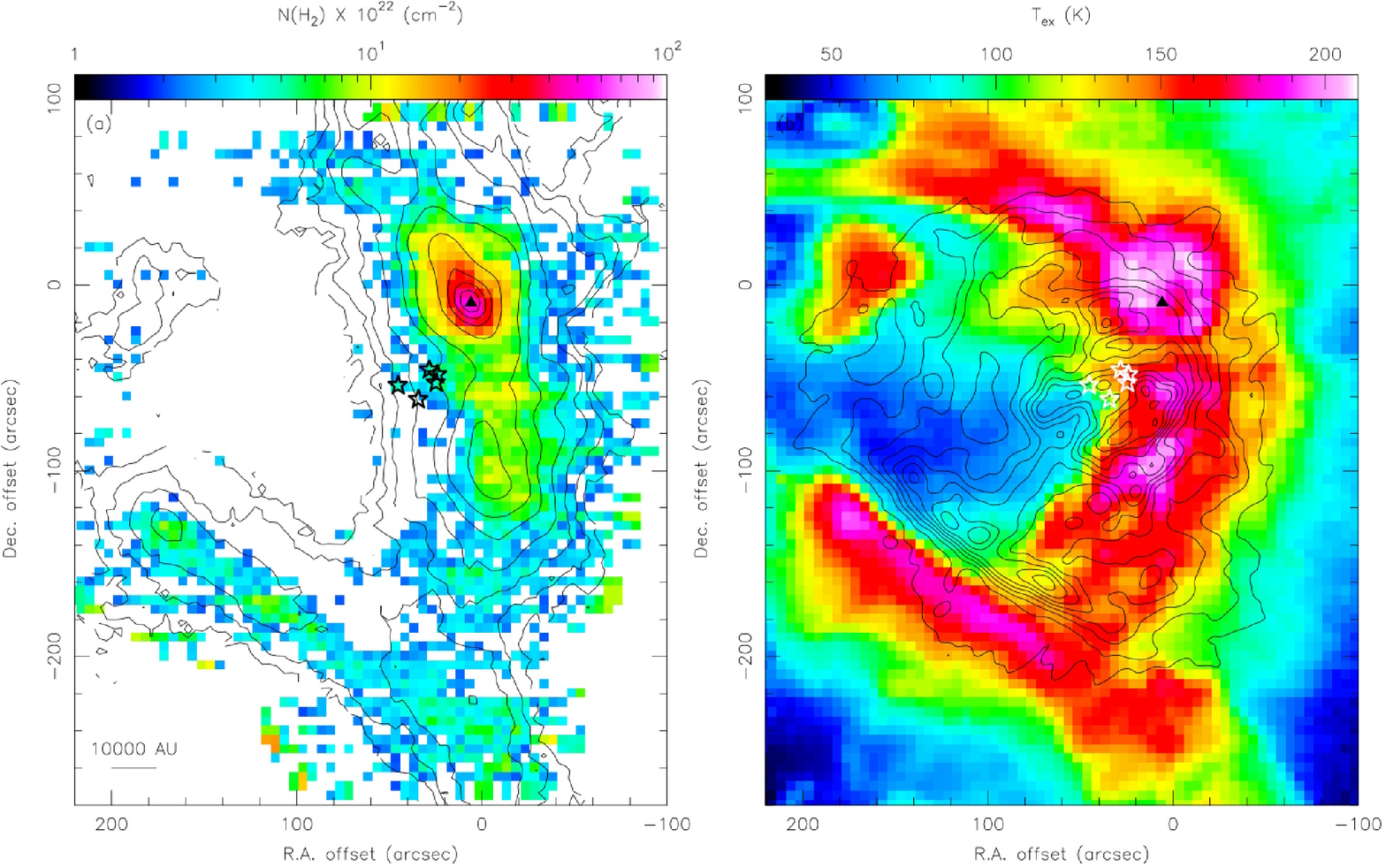}
 \caption{(a) The \CXVIIIO\ column density image of the OMC-1 core overlaid with black contours of the JCMT 850 \micron\ dust continuum image. The 850 \micron\ image is taken from the JCMT data archive. The contours represent 0.2$\%$, 0.5$\%$, 1$\%$, 2$\%$, 4$\%$, 7$\%$, 20$\%$, 40$\%$, 60$\%$, and 80$\%$ of the peak intensity (87 Jy beam$^{-1}$). (b) The excitation temperature image for \TWCO\ and \CXVIIIO\ overlaid with the 35 mm VLA-GBT continuum emission \citep{Dicker2009} in black contours running from 15$\%$ to 95$\%$ in steps of 8$\%$ of the peak intensity (0.94 Jy beam$^{-1}$). The stars mark the positions of the five Trapezium stars, and black triangles represent the position of the Orion Hot Core.\label{den-temp}}
 \end{figure*}

The results of the \HH\ column density and excitation temperature distributions are shown in Figure \ref{den-temp}, where the north-south dense ridge near the Trapezium cluster and the Orion Bar in the southeast are clearly seen. The average $\tau(\rm C^{18}O)$ at the $J=6-5$ transition over the whole OMC-1 core region is about 0.08 with an average excitation temperature of about 115 K, which are consistent with the \TWCO\ $J=9-8$ observations by \citet{Kawamura2002}. Besides, the minimum excitation temperature in the OMC-1 core is $\sim 30$ K, indicating a generally warm environment. The average \CXVIIIO\ column density in the OMC-1 core is about $8.7\times10^{15}$ cm$^{-2}$, corresponding to an \HH\ column density of $5.5\times10^{22}$ cm$^{-2}$ assuming a \TWCO\ abundance of $8\times10^{-5}$ \citep{Wilson1992}. Therefore, the dense gas mass is $\sim 140$ \Msol\ in an area of 0.13 pc$^2$ in the OMC-1 core. The \HH\ column density agrees with the density of $5\times10^{21}-7\times10^{22}$ cm$^{-2}$ derived from the mid-$J$ CO isotopologue observations by \citet{Wirstrom2006}. Our results agree with the recent measurement by \citet{Wilson2011} using the \THCO\ $J=6-5$ line with a lower [\TWCO]/[\HH] ratio of $2\times10^{-5}$. Additionally, \citet{Habart2010} derived an \HH\ column density of about $9\times10^{22}$ cm$^{-2}$ toward the Orion Bar from the {\it Herschel}/SPIRE high-$J$ CO observations, assuming a lower excitation temperature of 85 K and a higher \TWCO\ to \CXVIIIO\ ratio of 560. Their result is consistent with ours, where we obtained  an \HH\ column density of $\approx10^{23}$ cm$^{-2}$ toward the Orion Bar with a higher excitation temperature ($\gtrsim150$ K).

 \begin{figure*}
 \centering
 \includegraphics[angle=270,width=0.9\textwidth]{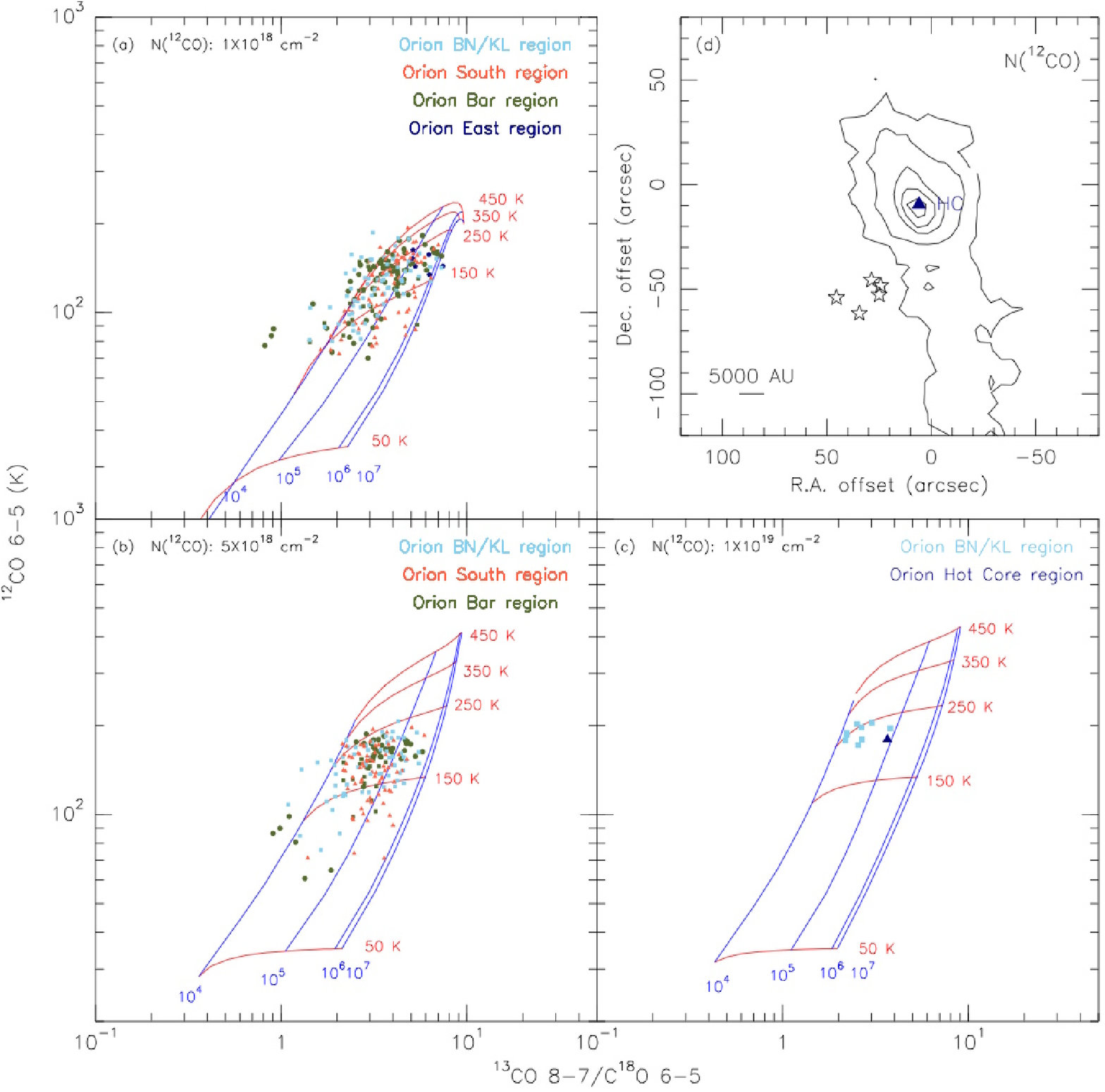}
 \caption{The RADEX modeling results shown in the \TWCO\ $J=6-5$ peak temperatures and the \THCO/\CXVIIIO\ ratios for three different \TWCO\ column densities. The column densities of \THCO\ and \CXVIIIO\ were fixed at the isotopologic abundance ratio of $1/60$ and $1/490$ of the \TWCO\ column density. The line widths of \TWCO, \THCO, and \CXVIIIO\ were fixed at 5, 5, and 4 \kms, respectively. The red and blue contours denote the temperature and \HH\ number density, respectively, and the color points represent the data in the different regions of the OMC-1 core. Each point represents a peak \TWCO\ $J=6-5$ temperature and a temperature ratio between \THCO\ $J=8-7$ and \CXVIIIO\ $J=6-5$ from a single pixel with a resolution of 20$\arcsec$. (a) The modeling input column density of \TWCO\ is $1\times10^{18}$ cm$^{-2}$, and the data are plotted in a \TWCO\ column density density of $7.5\times10^{17}-2.5\times10^{18}$ cm$^{-2}$. (b) The modeling input column density of \TWCO\ is $5\times10^{18}$ cm$^{-2}$, and the data are plotted in a density range of $2.5\times10^{18}-7.5\times10^{18}$ cm$^{-2}$. (c) The modeling input column density of \TWCO\ is $1\times10^{19}$ cm$^{-2}$, and the data are plotted in the density of $\geq7.5\times 10^{18}$ cm$^{-2}$. Upper right panel shows the \TWCO\ column density map toward Orion BN/KL in contours of 5, 10, 20, 30, and $40\times 10^{18}$ cm$^{-2}$. The stars mark the positions of the five Trapezium stars, and the Orion Hot Core (HC) position is also marked. \label{radex}}
 \end{figure*}

The JCMT 870 $\mu$m dust continuum emission image \citep{Johnstone1999} shown in Figure \ref{den-temp} agrees with our density map well, where the dust emission traces mostly the dense ridge (Orion BN/KL and South) and Bar regions. It is interesting to note that the dense ridge has an offset from the peak temperature positions, especially in the Orion South region ($\sim30\arcsec$) and the north of the Orion BN/KL region ($\sim20\arcsec$). This dense ridge seems to extend farther to the north of the OMC-1 core, and is probably related to the NH$_3$ filamentary structure seen by \citet{Wiseman1998,Wiseman1996}. Additionally, as Figure \ref{den-temp} (b) demonstrates, the interface between the ionized and the molecular warm dense gas is very pronounced, where Orion BN/KL and Orion South are located at the north-south dense ridge, and the Orion Bar and Orion East are part of the high temperature enclosed structure well illustrated by the free-free continuum emission. Furthermore, our large-scale images are consistent with the [\CII] and [\OI] emission maps obtained by \citet{Herrmann1997}, where the rather uniform [\CII] emission indicates that PDRs are present over the whole OMC-1 core region.

\subsection{RADEX modeling}

The non-LTE radiative transfer program RADEX \citep{vanderTak2007} was used to investigate the accuracy of the temperature and density calculation in LTE shown above. The optical depth effects are taken into account by RADEX using the escape probability approximation, where a uniform sphere geometry was chosen. The different transitions of \TWCO, \THCO, and \CXVIIIO\ were used in the modeling, and their line widths were fixed at 5, 5, and 4 \kms, respectively, which are the average values from our spectra. The molecular data using in the modeling are taken from the Leiden Atomic and Molecular Database\footnote{http://www.strw.leidenuniv.nl/$\sim$moldata/} \citep[LAMDA;][]{Schoeier2005}. In addition, only \HH\ was chosen as a collision partner, and 2.73 K was adopted as the background temperature. The input parameters are kinetic temperature and \HH\ number density. Each model was iterated with a kinetic temperature ranging from 50 to 450 K and an \HH\ number density ranging from $10^4$ to $10^8$ cm$^{-3}$. The input column densities of these three molecules were chosen, so that their ratios were fixed at their isotopologic abundance ratios, i.e., 490 for [\TWCO]/[\CXVIIIO] and 60 for [\TWCO]/[\THCO]. Three models were iterated with three different \TWCO\ column densities of $1\times10^{18}$, $5\times10^{18}$, and $1\times10^{19}$ cm$^{-2}$ which were obtained from the LTE calculation. Since these models did not take outflows into account, the output molecular line radiation temperatures ($T_{\rm R}$) were directly compared with the peak temperatures instead of the integrated intensities to avoid the strong line wing emission. The modeling results are shown in Figure \ref{radex}.

Figure \ref{radex} reveals that most regions in the OMC-1 core have an \HH\ number density of $\sim 10^{4}-10^{6}$ cm$^{-3}$. The Orion Bar and Orion East emission peaks (Fig. \ref{radex} a), however, show higher densities. The distribution of the \HH\ number density in the Bar is consistent with $10^{4}-10^{5}$ cm$^{-3}$ for a homogeneous medium \citep{Wyrowski1997}. The modeling results in the Orion BN/KL, South, and Bar regions all show similar trends of a diverse kinetic temperature (about 100--200 K), which are consistent with the LTE calculation results. In addition, the non-LTE modeling results indicate that the central region of Orion BN/KL close to the Hot Core (in 20\arcsec\ size) has a number density of about $10^5$ cm$^{-3}$ and a temperature of about 200 K, which is also close to the excitation temperature calculated in the LTE case. Noteworthily, the Orion BN/KL region and part of the Orion South region (Fig. \ref{radex} a) show very high kinetic temperatures ($>350$ K) which may indicate an extra heating mechanism, e.g., outflow/shock heating. The suggestion of outflow/shock heating is supported by the larger \TWCO\ line widths (Fig. \ref{map1}) and a great amount of known outflows in these regions \citep[see][]{Henney2007}. However, the high kinetic temperatures ($\geq250$ K) in the Orion Bar and Orion East regions (Fig. \ref{radex} a) are hardly explained by the outflow/shock heating for the lack of outflow activities or broad \TWCO\ line widths. Instead, as PDRs, the Orion Bar and Orion East are expected to be heated by FUV photons from the Trapezium OB stars. Moreover, a recent study of the Orion Bar \citep{Pellegrini2009} suggests an extra heating by the excess density of cosmic rays, which are trapped in the compressed magnetic field. 

\subsection{Mass of the OMC-1 core}

By adopting a radius of 40\arcsec--60\arcsec\ and an \HH\ density of $5\times10^{4}-2\times10^{5}$ cm$^{-3}$ for Orion BN/KL, we can estimate a gas mass of 7--85 \Msol\ with the assumption of a spherical volume at a distance of 414 pc \citep{Menten2007}. In a similar way, the gas mass in Orion South is 3--49 \Msol\ with a 30\arcsec--50\arcsec\ radius and an \HH\ density of $5\times10^{4}-2\times10^{5}$ cm$^{-3}$. For Orion East, the \HH\ density is $\sim2\times10^5$ cm$^{-3}$ with a 15\arcsec--25\arcsec\ radius, and the gas mass is 2--7 \Msol. In contrast, it is very difficult to estimate the gas mass in the Orion Bar because of its geometry. We adopted a cylindrical geometry and an \HH\ density of $5\times10^{4}-2\times10^{5}$ cm$^{-3}$ as the lower limit since an edge-on plane may contain much more gas. With a length of 300\arcsec\ and a radius of 20\arcsec--30\arcsec, the gas mass of the Orion Bar is 9--79 \Msol. The estimated total dense \HH\ gas mass in the OMC-1 core is 21--220 \Msol\ within a radius of 0.3 pc. This is also consistent with the mass estimated ($\sim140$ \Msol) from the LTE calculation above. If we include the warm diffuse gas ($10^{4}$ cm$^{-3}$) in the same volume, the total warm gas mass is 86--285 \Msol\ in the OMC-1 core. This  mass estimate for the OMC-1 core  is close to that derived from the Odin \TWCO\ and \THCO\ $J=5-4$ observations \citep{Wirstrom2006}, i.e., 320 \Msol\ in the molecular ridge for a farther distance of 500 pc. The result of \citet{Wilson2001} also suggests a similar estimate of a warm gas mass of 310--430 \Msol\ from the observations of \TWCO\ $J=7-6$ and $J=4-3$, again for a distance of 500 pc. Scaling these two literature values 
to the distance of 414 pc, results in 213--295 \Msol\ for the total mass of the OMC-1 core. This is comparable to the value derived by us. Hence, we conclude that the higher-$J$ CO lines do trace most of the gas in the OMC-1 core which is heated mainly by the Trapezium stars.



\section{Summary}

In this paper, we present the first large-scale images of highly excited \TWCO, \THCO, and \CXVIIIO\ lines in the OMC-1 core. The high excitation temperatures ($\sim$ 150--200
K) reveal a hot enclosed structure which is mostly heated by the Trapezium cluster. The LTE approximation and non-LTE RADEX program were used to derive the excitation
temperature and \HH\ density in the OMC-1 core. We found a typical density of $10^{4}-10^{6}$ cm$^{-3}$ in this region. The clear dense ridge in the north-south direction is
seen and is offset from the high temperature enclosed structure. Orion BN/KL and Orion South are located at this dense ridge, and the Orion Bar and East are part of the high
temperature enclosed structure. The estimated mass of total warm gas is 86--285 \Msol, assuming different geometries for different regions. In addition, the higher-$J$ CO lines trace most of the molecular gas in the OMC-1 core. The detailed investigation of individual region of the Orion BN/KL, Orion South, Orion Bar, and Orion East will be presented in the following papers.



\begin{acknowledgements}
We would like to thank the APEX staff for the support during the observations and Simon Dicker for providing the VLA-GBT map. This work was supported by the International Max Planck Research School (IMPRS) for Astronomy and Astrophysics at the Universities of Bonn and Cologne.
\end{acknowledgements}


\begin{thebibliography}{}

\bibitem[Bally et al.(2011)]{Bally2011} Bally, J., Cunningham, N.~J., Moeckel, N., et al.\ 2011, \apj, 727, 113 

\bibitem[Bally \& Zinnecker(2005)]{Bally2005} Bally, J., \&
    Zinnecker, H.\ 2005, \aj, 129, 2281


%

%
%
%
%
%
%
%
%
\bibitem[Bloemen et al.(1984)]{Bloemen1984} Bloemen, J.~B.~G.~M.,
    Caraveo, P.~A., Hermsen, W., Lebrun, F., Maddalena, R.~J., Strong, A.~W.,
    \& Thaddeus, P.\ 1984, \aap, 139, 37

 \bibitem[Boreiko
 \& Betz(1996)]{Boreiko1996} Boreiko, R.~T., \& Betz, A.~L.\ 1996, \apjl, 467, L113

%
%
\bibitem[Dame \& Thaddeus(1985)]{Dame1985} Dame, T.~M., \&
    Thaddeus,
    P.\ 1985, \apj, 297, 751

\bibitem[Dicker et al.(2009)]{Dicker2009} Dicker, S.~R., et al.\ 2009, 
\apj, 705, 226 

%
 \bibitem[Drapatz et
 al.(1983)]{Drapatz1983} Drapatz, S., Haser, L., Hofmann, R., Oda, N., \& Iyengar, K.~V.~K.\ 1983, \aap, 128, 207
%

%
%
%

    
\bibitem[Furuya 
\& Shinnaga(2009)]{Furuya2009} Furuya, R.~S., \& Shinnaga, H.\ 2009, \apj, 703, 1198

%
%
%
%
%
%
 \bibitem[Genzel
 \& Stutzki(1989)]{Genzel1989} Genzel, R., \& Stutzki, J.\ 1989, \araa, 27, 41
%
%
 \bibitem[Graf et al.(1990)]{Graf1990} Graf, U.~U., Genzel, R., Harris,
 A.~I., Hills, R.~E., Russell, A.~P.~G.,
 \& Stutzki, J.\ 1990, \apjl, 358, L49
%
%
%
%
%
%

\bibitem[Goldsmith et al.(1997)]{Goldsmith1997} Goldsmith, P.~F., Bergin, E.~A., \& Lis, D.~C.\ 1997, \apj, 491, 615 



\bibitem[G{\"u}sten et al.(2006)]{Gusten2006} G{\"u}sten, R., Nyman, 
L.~{\AA}., Schilke, P., Menten, K., Cesarsky, C., 
\& Booth, R.\ 2006, \aap, 454, L13 

%
\bibitem[G\"usten et al.(2008)]{Gusten2008} G\"usten, R., Baryshev, A., Bell, A., Belloche, A., Graf, U., Hafok, H., Heyminck, S., et al.\ 2008,
\procspie, 7020, 25
%

\bibitem[Habart et al.(2010)]{Habart2010} Habart, E., et al.\ 2010, \aap, 
518, L116 



 \bibitem[Henney et al.(2007)]{Henney2007} Henney, W.~J., O'Dell,
 C.~R., Zapata, L.~A., Garc{\'{\i}}a-D{\'{\i}}az, M.~T., Rodr{\'{\i}}guez,
 L.~F., \& Robberto, M.\ 2007, \aj, 133, 2192
%
 \bibitem[Herrmann et al.(1997)]{Herrmann1997} Herrmann, F., Madden, S.~C.,
 Nikola, T., Poglitsch, A., Timmermann, R., Geis, N., Townes, C.~H.,
 \& Stacey, G.~J.\ 1997, \apj, 481, 343
%
%
 \bibitem[Hollenbach
 \& Tielens(1997)]{Hollenbach1997} Hollenbach, D.~J., \& Tielens, A.~G.~G.~M.\ 1997, \araa, 35, 179
%
%
 \bibitem[Houde et al.(2004)]{Houde2004} Houde, M., Dowell, C.~D.,
 Hildebrand, R.~H., Dotson, J.~L., Vaillancourt, J.~E., Phillips, T.~G.,
 Peng, R., \& Bastien, P.\ 2004, \apj, 604, 717
%
%
\bibitem[Johnstone 
\& Bally(1999)]{Johnstone1999} Johnstone, D., \& Bally, J.\ 1999, \apjl, 510, L49 


%
\bibitem[Kasemann et al.(2006)]{Kasemann2006} Kasemann, C., et al.\ 2006,
 \procspie, 6275

\bibitem[Kawamura et
al.(2002)]{Kawamura2002} Kawamura, J., et al.\ 2002, \aap, 394, 271

    




\bibitem[Klein et al.(2006)]{Klein2006} Klein, B., Philipp, S.~D.,
Kr{\"a}mer, I., Kasemann, C., G{\"u}sten, R.,
\& Menten, K.~M.\ 2006, \aap, 454, L29

\bibitem[Kurtz et al.(2000)]{Kurtz2000} Kurtz, S., Cesaroni, R.,
    Churchwell, E., Hofner, P., \& Walmsley, C.~M.\ 2000, Protostars and
    Planets IV, 299




%
%
\bibitem[Kutner et al.(1976)]{Kutner1976} Kutner, M.~L., Evans, N.~J., II, \&
    Tucker, K.~D.\ 1976, \apj, 209, 452
%




\bibitem[Lis et al.(1998)]{Lis1998} Lis, D.~C., Serabyn, E., Keene, J.,
Dowell, C.~D., Benford, D.~J., Phillips, T.~G., Hunter, T.~R.,
\& Wang, N.\ 1998, \apj, 509, 299


\bibitem[Marrone et al.(2004)]{Marrone2004} Marrone, D.~P., et al.\ 2004,
\apj, 612, 940

\bibitem[Masson et al.(1984)]{Masson1984} Masson, C.~R., et al.\ 1984,
    \apjl, 283, L37


\bibitem[Menten et al.(2007)]{Menten2007} Menten, K.~M., Reid, M.~J., Forbrich, J., \& Brunthaler, A.\ 2007, \aap, 474, 515

%



\bibitem[Mezger et al.(1990)]{Mezger1990} Mezger, P.~G., Zylka, R., \& Wink, J.~E.\ 1990, \aap, 228, 95



%

\bibitem[M{\" u}ller et al.(2005)]{Mueller2005} M{\" u}ller, H. S. P., Schl{\" o}der, F., Stutzki, J., \& and Winnewisser, G. 2005
J. Mol. Struct. 742, 215

\bibitem[O'Dell(2001)]{O'Dell2001} O'Dell, C.~R.\ 2001, \araa, 39, 99

\bibitem[O'Dell et al.(2008)]{O'Dell2008} O'Dell, C.~R., Muench,
A., Smith, N.,
\& Zapata, L.\ 2008, Handbook of Star Forming Regions, Volume I:
The Northern Sky ASP Monograph Publications, Vol.~4.~Edited by Bo Reipurth, p.544, 544



\bibitem[Pellegrini et al.(2009)]{Pellegrini2009} Pellegrini, E.~W.,
Baldwin, J.~A., Ferland, G.~J., Shaw, G.,
\& Heathcote, S.\ 2009, \apj, 693, 285

%
%
%
%
 \bibitem[Risacher et al.(2006)]{Risacher2006} Risacher, C., et al.\ 2006,
 \aap, 454, L17
%
%
%
%
%

 \bibitem[Schmid-Burgk et al.(1989)]{Schmid-Burgk1989} Schmid-Burgk, J., et
 al.\ 1989, \aap, 215, 150
 
 \bibitem[Sch{\"o}ier et al.(2005)]{Schoeier2005} Sch{\"o}ier, F.~L., van der Tak, F.~F.~S., van Dishoeck, E.~F., \& Black, J.~H.\ 2005, \aap, 432, 369
%
%
%
%
%
%
%
%

%



\bibitem[van der Tak et al.(2007)]{vanderTak2007} van der Tak, F.~F.~S., 
Black, J.~H., Sch{\"o}ier, F.~L., Jansen, D.~J., 
\& van Dishoeck, E.~F.\ 2007, \aap, 468, 627 

%
 \bibitem[Walmsley et al.(2000)]{Walmsley2000} Walmsley, C.~M., Natta, A., Oliva, E., \& Testi, L.\ 2000, \aap, 364, 301
%
\bibitem[Wilner et al.(1994)]{Wilner1994} Wilner, D.~J., Wright,
M.~C.~H., \& Plambeck, R.~L.\ 1994, \apj, 422, 642



%
 \bibitem[Wilson
 \& Matteucci(1992)]{Wilson1992} Wilson, T.~L., \& Matteucci, F.\ 1992, \aapr, 4, 1
%
%
 \bibitem[Wilson et al.(2001)]{Wilson2001} Wilson, T.~L., Muders, D., Kramer, C., \& Henkel, C.\ 2001, \apj, 557, 240
%
\bibitem[Wilson et al.(2011)]{Wilson2011} Wilson, T.~L., Muders, D., Dumke, 
M., Henkel, C., \& Kawamura, J.~H.\ 2011, \apj, 728, 61 




%
 \bibitem[Wirstr{\"o}m et al.(2006)]{Wirstrom2006} Wirstr{\"o}m, E.~S.,
 Bergman, P., Olofsson, A.~O.~H., Frisk, U., Hjalmarson, {\AA}., Olberg, M.,
 Persson, C.~M., \& Sandqvist, A.\ 2006, \aap, 453, 979
%
%

 \bibitem[Wiseman
 \& Ho(1996)]{Wiseman1996} Wiseman, J.~J., \& Ho, P.~T.~P.\ 1996, \nat, 382, 139

 \bibitem[Wiseman
 \& Ho(1998)]{Wiseman1998} Wiseman, J.~J., \& Ho, P.~T.~P.\ 1998, \apj, 502, 676
%
%

%
%

\bibitem[Wyrowski et al.(1997)]{Wyrowski1997} Wyrowski, F., Schilke, P., Hofner, P., \& Walmsley, C.~M.\ 1997, \apjl, 487, L171

\bibitem[Zapata et al.(2005)]{Zapata2005} Zapata, L.~A., Rodr{\'{\i}}guez,
L.~F., Ho, P.~T.~P., Zhang, Q., Qi, C.,
\& Kurtz, S.~E.\ 2005, \apjl, 630, L85

\bibitem[Zapata et al.(2006)]{Zapata2006} Zapata, L.~A., Ho,
P.~T.~P., Rodr{\'{\i}}guez, L.~F., O'Dell, C.~R., Zhang, Q.,
\& Muench, A.\ 2006, \apj, 653, 398

\bibitem[Zapata et al.(2009)]{Zapata2009} Zapata, L.~A., Schmid-Burgk, J., 
Ho, P.~T.~P., Rodr{\'{\i}}guez, L.~F., 
\& Menten, K.~M.\ 2009, \apjl, 704, L45 

\bibitem[Zapata et al.(2010)]{Zapata2010} Zapata, L.~A., Schmid-Burgk, J., 
Muders, D., Schilke, P., Menten, K., \& Guesten, R.\ 2010, \aap, 510, A2

\bibitem[Zapata et al.(2011)]{Zapata2011} Zapata, L.~A., Schmid-Burgk, J., \& Menten, K.~M.\ 2011, \aap, 529, A24 


\bibitem[{{Zuckerman}(1973)}]{Zuckerman1973} {Zuckerman}, B. 1973, \apj, 183,
    863



\end{thebibliography}
\end{document}